\documentclass[conference, twocolumn, 9pt]{IEEEtran}	

\PassOptionsToPackage{utf8}{inputenc}
\usepackage{inputenc}
\usepackage{cite}
\PassOptionsToPackage{T1}{fontenc} 
\usepackage{fontenc}
\PassOptionsToPackage{dvipsnames}{xcolor}
\usepackage{soul} 

\PassOptionsToPackage{fleqn}{amsmath}       
\usepackage{amsmath}
\usepackage{mathtools}
\usepackage{xfrac}
\usepackage{siunitx}
\usepackage[inline]{enumitem}

\usepackage{balance}

\usepackage[vlined,ruled,linesnumbered,longend]{algorithm2e}
\usepackage{graphicx}
\usepackage{textcomp}
\usepackage{xspace}
\usepackage[size=tiny]{todonotes}
\usepackage[hidelinks]{hyperref}

\usepackage{subfig}
\usepackage{multirow}
\usepackage{multicol}
\usepackage[export]{adjustbox}
\usepackage{float}
\usepackage{booktabs}
\usepackage[super]{nth}
\usepackage{amsthm}

\usepackage{bm} 

\newcommand{\ie}{i.\,e.\xspace}

\newcommand{\fiction}{\emph{fiction}\xspace}

\newcommand{\quicksim}{\emph{\mbox{QuickSim}}\xspace}

\newcommand{\quickexact}{\emph{\mbox{QuickExact}}\xspace}

\newcommand{\exgs}{\emph{\mbox{ExGS}}\xspace}

\newcommand{\simanneal}{\emph{\mbox{SimAnneal}}\xspace}

\newcommand{\hsi}{\mbox{H-Si(100)-2\ensuremath{\times}1}\xspace}

\theoremstyle{definition}

\newcommand{\refsec}[1]{Section~\ref{sec:#1}}
\newcommand{\reffig}[1]{Fig.~\ref{fig:#1}}

\newcommand{\reftab}[1]{Table~\ref{tab:#1}}

\newcommand{\refalgo}[1]{Alg.~\ref{algo:#1}}
\newcommand{\refline}[1]{Line~\ref{line:#1}}

\newcommand{\refequa}[1]{Eq.~(\ref{eq:#1})}

\newtheorem{example}{Example}

\title{ \huge The Need for Speed:  \\ Efficient Exact Simulation of Silicon Dangling Bond Logic}

\author{\IEEEauthorblockN{Jan Drewniok\IEEEauthorrefmark{1}, Marcel Walter\IEEEauthorrefmark{1}, and Robert Wille\IEEEauthorrefmark{1}\IEEEauthorrefmark{3}} \\[-1ex]
		\IEEEauthorblockA{\IEEEauthorrefmark{1}Chair for Design Automation, Technical University of Munich, Germany \\
				\IEEEauthorblockA{\IEEEauthorrefmark{3}Software Competence Center Hagenberg GmbH, Austria}} \\[-2ex]
		Email: \{\href{mailto:jan.drewniok@tum.de}{jan.drewniok}, \href{mailto:marcel.walter@tum.de}{marcel.walter}, \href{mailto:robert.wille@tum.de}{robert.wille}\}@tum.de}

\begin{document}
	
	\maketitle
	\thispagestyle{empty}
	\pagestyle{empty}

	\begin{abstract}
		\emph{The Silicon Dangling Bond}~(SiDB) logic platform, an emerging computational \mbox{beyond-CMOS} nanotechnology, is a promising competitor due to its ability to achieve integration density and clock speed values that are several orders of magnitude higher compared to current CMOS fabrication nodes. However, the exact physical simulation of SiDB layouts, which is an essential component of any design validation workflow, is computationally expensive. In this paper, we propose a novel algorithm called \quickexact, which aims to be both, efficient and exact. To this end, we are introducing three techniques, namely \begin{enumerate*}
			\item \mbox{Physically-informed} Search Space Pruning,
			\item Partial Solution Caching, and
			\item Effective State Enumeration
		\end{enumerate*}.
		Extensive experimental evaluations confirm that, compared to the \mbox{state-of-the-art} algorithm, the resulting approach leads to a paramount runtime advantage of more than a factor of \SI{5000}{} on randomly generated layouts and more than a factor of \SI{2000}{} on an established gate library. 
	\end{abstract}
	
	
	\section{Introduction \& Motivation}

	
	As the limits of Moore's Law are approached and processing technologies reach the top of the \mbox{S-curve}, there is a growing interest in the \emph{Silicon Dangling Bond}~(SiDB) logic platform, an emerging computational \mbox{beyond-CMOS} nanotechnology~\cite{D0NR08295C, wyrick2019atom, haider2009controlled, huff2017atomic, pavlicek2017tip, achal2018lithography, Wolkow14}. One of the key features of SiDB logic are its \mbox{sub-nanometer} elementary devices, which can significantly improve integration density by several orders of magnitude compared to current CMOS fabrication nodes~\cite{huff2018binary, huff2017atomic, achal2018lithography, wang2020atomic, haider2009controlled, Wolkow14, pitters2011charge, rashidi2018initiating}. With contemporary \mbox{\emph{transistor-based}} technology requiring vast amounts of energy and being limited to low GHz speeds, SiDB technology reduces energy consumption for similar processes by over \SI{99}{\percent} while enabling an increase in processing speeds of several orders of magnitude \cite{livadaru2010danglingbond}. This makes SiDB logic promising for achieving \mbox{ultra-low} energy dissipation and establishes it as a \mbox{much-anticipated} green competitor in the \mbox{beyond-CMOS} domain~\mbox{\cite{landauer1961irreversibility, keyes1970minimal, toth1999quasiadiabatic, huff2018binary, rashidi2018initiating, haider2009controlled}}. In addition, SiDB logic has been proposed as a candidate for the integration of quantum computers with conventional CMOS circuitry~\cite{Wolkow14, dzurak2001construction}.
	
	The growing interest in the SiDB logic platform has led to proposals from the scientific community for gate and circuit libraries~\cite{walter2022hexagons, ng2020siqad, vieira2022three, bahar2020atomic} and design automation solutions~\cite{ng2020siqad, walter2022hexagons, lupoiu2022automated, Hofmann2023nanoplacer, hofmann202345degreeturn}. Furthermore, the technology is gaining commercial momentum, as evidenced by the recently formed research enterprise \emph{Quantum Silicon Inc.}, which aims to be one of the first industry adopters of SiDBs and has secured \mbox{multi-million dollar} investments~\cite{Wolkow2021, Wolkow2021a}. These developments are also putting pressure on design and simulation tools to keep pace, especially with the rapid advances in SiDB fabrication capabilities~\cite{haider2009controlled, huff2017atomic, pavlicek2017tip, achal2018lithography}. In particular, physical simulation is an essential component of any design validation workflow, providing the basis for verifying the behavior of circuit layouts before fabrication.
	
	However, physical simulation of SiDBs is challenging: To accurately predict the behavior of a system of~$n$~SiDBs, it is necessary to solve a \mbox{high-dimensional} optimization problem that takes into account all relevant physical effects. This involves enumerating up to $3^n$ charge configurations and checking their physical validity. Due to this exponential runtime complexity, such an algorithm is limited to small instances.
	
	In the literature, two approximate algorithms for solving this problem in polynomial time were proposed, namely \quicksim~\cite{drewniok2023quicksimIEEE} and \simanneal~\cite{ng2020siqad}.
	In contrast, only one exact algorithm that determines \emph{all} physical valid charge configurations 
	has been presented: \mbox{\emph{ExhaustiveGS}~(ExGS)~\cite{ng2020thes}}. It performs an exhaustive search of the entire (exponential) search space. Therefore, \exgs is severely limited in its runtime efficiency and, thus, is not even applicable to \mbox{medium-sized} SiDB layouts like standard gates. This urgently calls for a more sophisticated and efficient exact simulation approach for SiDBs.
	
	To this end, we propose a novel exact simulation algorithm denoted \quickexact, which is designed with a special focus on runtime efficiency. Our approach utilizes three main techniques, namely \begin{enumerate*}
		\item \mbox{Physically-informed} Search Space Pruning,
		\item Partial Solution Caching, and
		\item Effective State Enumeration
	\end{enumerate*}. Extensive experimental evaluations, that are covering the simulation of established gate libraries as well as random instances, confirm that the resulting approach leads to a paramount runtime advantage of more than a factor of \SI{5000}{} on randomly generated layouts and more than a factor of \SI{2000}{} on an established gate library compared to the \mbox{state-of-the-art} algorithm \exgs. 
	
	To establish this paper as a \mbox{self-contained} work, \refsec{prelims} provides an overview of the basic concepts of SiDB systems and their physical simulation. Afterwards, related work from the literature is discussed in \refsec{simulation:approaches}. \refsec{quickexact} constitutes the main contribution of this work as it provides a detailed introduction to the proposed algorithm \quickexact. \refsec{eval} presents exhaustive experimental evaluations demonstrating the advantages of \quickexact over \mbox{state-of-the-art} techniques. Finally, \refsec{concl} concludes the paper.
	
	\section{Preliminaries} \label{sec:prelims}
	
	In this section, preliminaries are established that are necessary for the comprehension of the remainder of this work. First, an overview of the SiDB logic platform is given in \refsec{prelims:sidbs}, followed by a comprehensive presentation of the physical simulation of SiDBs in \refsec{prelims:simulation}.
	
	\subsection{The SiDB Logic Platform} \label{sec:prelims:sidbs}
	
\begin{figure}[t!]
		\centering
		\subfloat[SiDB fabrication on the \hsi surface (side view)]{
			\quad\includegraphics[width=.35\linewidth]{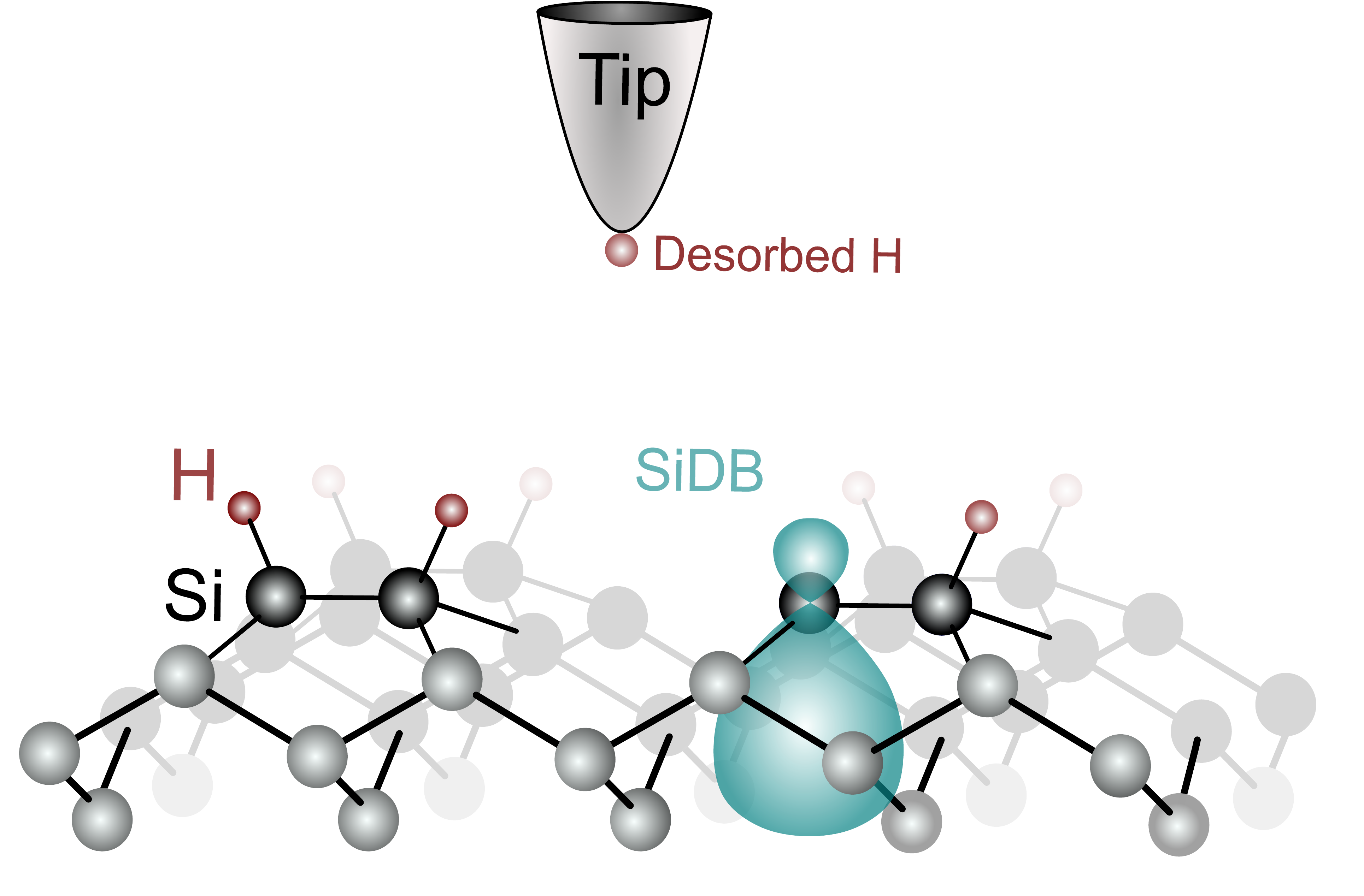}\quad
			\label{fig:sidbs:generation}
		} \hfil
		\subfloat[Surface lattice top view]{
			\includegraphics[width=.35\linewidth]{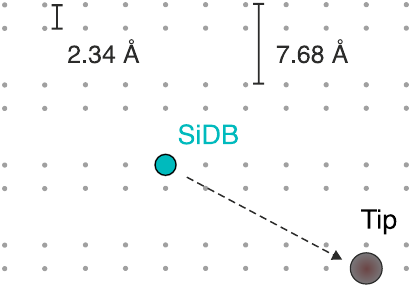}
			\label{fig:sidbs:lattice}
		} \hfil
		\\
		\hfil
		\subfloat[Energy landscape of three interacting SiDBs]{
			\includegraphics[width=.9\linewidth]{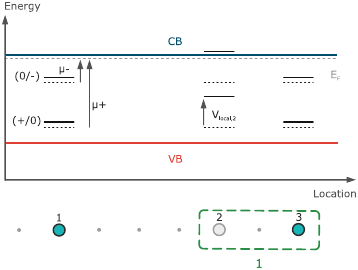}
			\label{fig:sidbs:energy_level}
		} \hfil
		\subfloat[BDL wire segment transmitting a binary \texttt{1}]{
			\includegraphics[width=.9\linewidth]{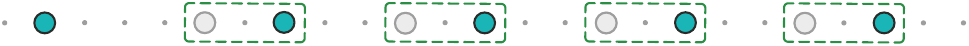}
			\label{fig:sidbs:wire}
		} \\
		\subfloat[OR-gate (\texttt{10})~\cite{huff2018binary, ng2020siqad}]{
			\includegraphics[width=.4\linewidth]{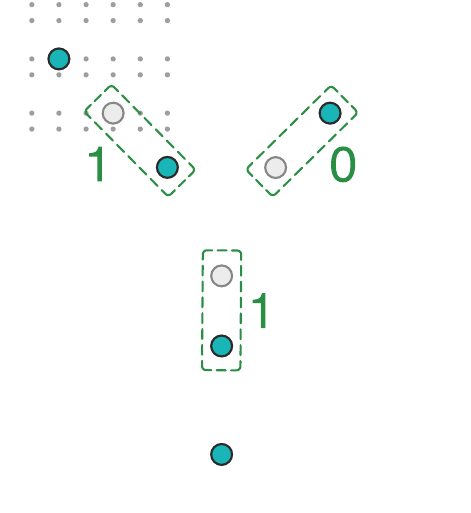}
			\label{fig:sidbs:or}
		} \hfil
		\subfloat[AND-gate (\texttt{10})~\cite{ng2020siqad}]{
			\includegraphics[width=.4\linewidth]{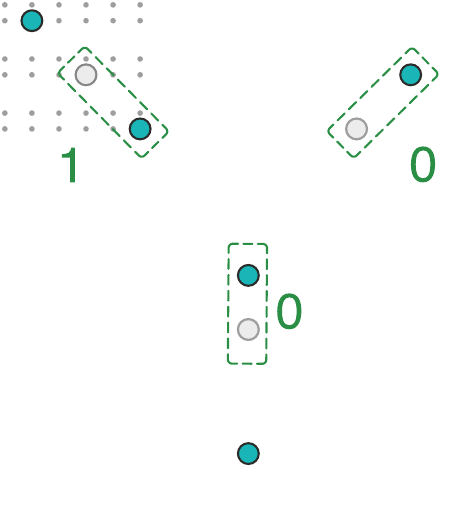}
			\label{fig:sidbs:and}
		} \\
		\caption{Elemental SiDB logic components}
		\label{fig:sidbs}
	\end{figure}

	\emph{Silicon Dangling Bonds}~(SiDBs) are created on a \emph{Hydrogen-passivated Silicon} surface~(\hsi) by utilizing a \emph{Scanning Tunneling Microscope}~(STM) with an atomically sharp tip~\cite{pavlicek2017tip, huff2017atomic, achal2018lithography, rashidi2022automated, huff2019landscape, achal2019detecting}. Applying a voltage through the STM disrupts the covalent bond between a silicon and a hydrogen atom, causing the hydrogen to be removed from the sample and desorbed to the tip, leaving behind a \emph{dangling} valence bond. The process is sketched in \reffig{sidbs:generation} and a top view of the surface lattice is shown in \reffig{sidbs:lattice}, where the teal dot indicates the SiDB. The tip is then moved to a new silicon dimer and the process is repeated to create another SiDB. Each SiDB acts as a chemically identical quantum dot that can trap a maximum of two electrons in its \mbox{$sp^{3}$-orbital}~\cite{taucer2015silicon}. Therefore, each SiDB is confined to one of three different states: negatively, neutrally, or positively charged.  
	
	The charge state of an SiDB depends on the
	\mbox{charge-transition} energy levels, which are shown in \reffig{sidbs:energy_level} for three SiDBs. When an SiDB is undisturbed (\ie, no other SiDBs or external electrostatic potentials are present), the charge transition levels (\mbox{$0$\,/\,$-$}) and (\mbox{$0$\,/\,$+$}) (dashed lines) lie roughly $\SI{0.3}{\eV}$ and $\SI{0.8}{\eV}$ below the conduction band~(CB)~\cite{rashidi2016timeresolved, pitters2011charge}. The location of the energy transition levels with respect to the \emph{Fermi energy}~$E_{F}$ defines the charge state of the SiDB. If \mbox{($0$\,/\,$-$)} is below~$E_{F}$, the SiDB is negatively charged. Vice versa, if only \mbox{($+$\,/\,$0$)} is below~$E_{F}$, the SiDB is neutrally charged, and in the last case, the SiDB contains no more electrons and is therefore positively charged. In \mbox{n-doped} systems, $E_{F}$ is close to the CB and thus, SiDBs are negatively charged when not disturbed. However, if several SiDBs are placed close together as illustrated in \reffig{sidbs:energy_level}, the single charge transition levels can be shifted with respect to~$E_{F}$ by electrostatic potential interaction among SiDBs. 
	As illustrated, the two outermost SiDBs generate a local electrostatic potential at the second SiDB of $V_{\mathit{local},2} > \SI{0.4}{\eV}$, which shifts \mbox{($0$\,/\,$-$)} above the Fermi energy, which then leads to a neutrally charged second SiDB. This yields a \emph{\mbox{Binary-Dot Logic}}~(BDL) pair with a discrete charge state realizing the binary state $1$ in this case, as illustrated by the green rectangle. This specific property is exploited to utilize SiDBs for the fabrication of logic devices~\cite{huff2018binary}. 
	
	A sequence of BDL pairs, depicted in \reffig{sidbs:wire}, can operate as a binary wire that transmits information solely through the coupling of electric fields. This principle can also be employed to create an \mbox{OR-gate}, as demonstrated in \reffig{sidbs:or}, utilizing the binary input \texttt{10}, where the repulsion (electrostatic coupling) leads to an output of \texttt{1}. In fact, both a BDL wire consisting of eight SiDBs and an \mbox{OR-gate} (with a total size smaller than \SI{30}{\square\nano\meter}) have already been realized experimentally by Huff~et~al.~\cite{huff2018binary}. Additionally, this concept can be extended to implement other gates, such as the \mbox{AND-gate} depicted in \reffig{sidbs:and}, whose BDL pairs are differently spaced.

	\subsection{Physical Simulation} \label{sec:prelims:simulation}
	
	In order to validate the designs of SiDB assemblies and facilitate the advancement of this technology, the utilization of physical simulation is deemed imperative. The objective of such simulation is to accurately simulate the charge distribution of a given arrangement of SiDBs. To this end, electrostatic potential simulation is conducted. The electrostatic potential $V_{i,j}$ at \mbox{position $i$} generated by an SiDB in the state~\mbox{$n_{j} \in \{-1,0,1\}$} at \mbox{position $j$} is given by~\cite{huff2018binary, huff2019landscape}
	\begin{align}
		V_{i,j} = -\frac{q_{e}}{4  \pi \epsilon_{0} \epsilon_{r}} \cdot \frac{e^{- \frac{{d_{i,j}}}{\lambda_\mathit{tf}}}}{d_{i,j}} \cdot n_{j},
		\label{eq:potential_sidb}
	\end{align}
	where $\lambda_{\mathit{tf}}$ defines the \emph{Thomas-Fermi screening length} and $\epsilon_{r}$ the \emph{dielectric constant}, which were experimentally extracted to be \SI{5}{\nm} and $5.6$, respectively~\cite{huff2018binary}. Moreover, $\epsilon_{0}$, $q_{e}$, and $d_{i,j}$ are the \emph{vacuum permittivity}, the \emph{electron charge ($q_{e}=-e$; $e$: elementary charge)}, and the \emph{Euclidean distance} between position $i$ and $j$, respectively. The layout's total electrostatic potential \mbox{energy $E$} is thus given as
	\begin{align}
		E =  -\sum_{i<j} V_{i,j} \cdot n_{i} \cdot q_{e}.
		\label{eq:energy}
	\end{align}
	The expected state of an SiDB layout is determined by the charge configuration with the lowest energy. Therefore, in order to determine said state, \refequa{energy} has to be minimized. However, a solution (\ie, a valid charge distribution) has to meet the physical constraint of \emph{metastability}, which can be described by the combination of two criteria, namely \emph{population stability} and \emph{configuration stability}. Both must be obeyed to guarantee simulation result validity.
	
	\paragraph{Population Stability} 
	\label{population_stability}
	As highlighted before in \reffig{sidbs:energy_level}, the charge state of each SiDB must be consistent with the energy \mbox{charge-transition} levels \mbox{($0$\,/\,$-$)} and \mbox{($+$\,/\,$0$)} relative to~$E_{F}$. The energy of the SiDB's charge transition level at position~$i$ depends on the local electrostatic potential $V_{\mathit{local},i}$ at that position defined by
	\begin{align}
		V_{local,i} = \sum_{j,\, j \neq i} V_{i,j}. 
		\label{eq:local_potential}
	\end{align}
	Intuitively, this means that an SiDB is preferably neutrally charged when adjacent SiDBs are negatively charged and vice versa. This relation is formally expressed by the following conditions: SiDB- when $\mu_{-} + V_{\mathit{local},i} \cdot q_{e} < 0$, SiDB+ when $ \mu_{+} + V_{\mathit{local},i} \cdot q_{e} > 0 $, and SiDB0 otherwise. 
	
	\paragraph{Configuration Stability} 	\label{configuration_stability}
	If there is no feasible \mbox{single-electron} hop event between arbitrary SiDBs leading to a lowering of the system's total electrostatic potential energy, the charge state fulfills the configuration stability. In other words, if this constraint is not satisfied, there would exist a state of lower electrostatic potential energy that the system would spontaneously converge to. 
	
%

Overall, the charge configuration with the lowest electrostatic potential system energy that is also physically valid, \ie, satisfies metastability, is called the system's \emph{ground state}. It represents the charge configuration of a given SiDB system at low temperature and, thereby, its physical behavior~\cite{drewniok2023temperatureIEEE}. 
	
	\section{Related Work} \label{sec:simulation:approaches}
	In this section, the exact SiDB simulator \emph{ExhaustiveGS}~(\exgs)~\cite{ng2020thes} from the literature is discussed.
	As described above, SiDBs can be present in one of three different charge states. Since their charge state depends on the location of the energy transition level with respect to $E_{F}$, which is in turn defined by the charge states of the surrounding SiDBs~(cf.~\reffig{sidbs:energy_level}), a \mbox{high-dimensional} interwoven problem has to be solved in order to determine the ground state and all other physically valid charge configurations 
	of the system. 
	
	Mastering the complexity of simulation is a critical aspect of any design automation process for \mbox{SiDB-based} logic. 
	However, \emph{exact} physical simulation of SiDBs is a computationally complex task for which there is only \exgs~\cite{ng2020thes}. 
	
	\exgs enumerates all possible charge configurations of a given SiDB layout, computes the local electrostatic potential and the system's electrostatic potential energy with \refequa{energy} and \refequa{local_potential} from scratch for each single charge configuration. At the same time, it determines whether each charge configuration is physically valid. This eventually allows obtaining not only the ground state, but all physically valid charge configurations of a given SiDB layout.
	Since \exgs considers \emph{all} possible charge configurations (and does not apply search space pruning or similar methods to reduce the resulting computational complexity), already small gate layouts cannot be simulated in a reasonable amount of time.  This shortcoming is caused by the inherent exponential complexity of the SiDB simulation task.
	\begin{example}
		In order to simulate a layout consisting of $32$~SiDBs in \mbox{3-state} simulation (SiDBs are also allowed to be positively charged), a total of $3^{32} \approx 10^{15}$ charge configurations have to be considered and checked for physical validity. 
		
	\end{example}
	
	
	
	Ultimately, an algorithm that is both efficient and exact (\ie, able to find all physically valid charge distributions) is desired. This algorithm is presented in this paper, with the general idea as well as the details elaborated in the next section.

	\section{Proposed Algorithm: \quickexact} \label{sec:quickexact}
	In this section, the proposed SiDB simulation algorithm \quickexact is introduced. First, the general idea is outlined in \refsec{quickexact:idea} and the techniques that lead to the performance advantage of more than three orders of magnitude compared to state of the art are elaborated, with their precise impact analyzed in \refsec{quickexact:complexity}.
	Finally, the implementation details of \quickexact are presented in \refsec{quickexact:implementation}.
	
	\subsection{General Idea} \label{sec:quickexact:idea}
	To reiterate, \exgs is the \mbox{state-of-the-art} \emph{exact} physical SiDB simulator. However, it is highly inefficient and is, thus, not applicable to even simulate medium-sized SiDB assemblies like gates. Therefore, there exists a \emph{Need for Speed}.
	To simulate efficiently while being exact at the same time, we propose the three techniques: 
	\begin{enumerate}
		\item Physically-informed Seach Space Pruning,
		\item Partial Solution Caching, and
		\item Effective State Enumeration.
	\end{enumerate}
	The following segments elaborate on their respective ideas.

	\subsubsection{\mbox{Physically-informed} Search Space Pruning} \label{sec:quickexact:pruning}
	If the charge state of $d$ SiDBs (in a physically valid charge distribution) is known prior to simulation, the search space is reduced by a factor of~$2^{d}$ or~$3^{d}$ for \mbox{2-state} and \mbox{3-state} simulation, respectively. In fact, SiDBs that only weakly interact with other ones, have a tendency to be always negatively charged under realistic $\mu_{-}$ values. 
	With respect to the population stability discussed in \refsec{prelims:simulation}, this can be expressed as the following relation: the \mbox{$i$-th} SiDB has to be negatively charged for all physically valid charge distributions if the maximal possible local electrostatic potential $V_{\mathit{local, max},i}$ at position~$i$ is smaller than $\mu_{-}$. This is summarized by the following constraint:
	
	\begin{align}\label{eq:negative_sidbs}
		\mathit{SiDB_{i,-}} \iff \mu_{-} < -V^{\mathit{max}}_{\mathit{local},i} \cdot q_{e}
	\end{align}

	According to \refequa{local_potential}, $V_{\mathit{local},i}$ is maximized if each $V_{i,j}$ is positive. Due to \refequa{potential_sidb}, this is the case if $\forall i: n_{i} = -1$. Ultimately, in order to detect all SiDBs that have to be negatively charged to fulfill the metastability, all SiDBs are set to negative. Afterwards, the SiDBs that fulfill \refequa{negative_sidbs} are determined. This whole process is sketched by the pseudocode in \refalgo{negative_sidbs}, where all SiDBs are set to be negatively charged in \refline{negative_sidbs:negative_distribution}. Afterwards, all SiDBs that fulfill \refequa{negative_sidbs} (\refline{negative_sidbs:detected}) are collected and returned (\refline{negative_sidbs:return}).
	
	To reduce the search space further by an additional factor of $2$ (or $3$ for \mbox{3-state} simulation), a \mbox{so-called} \emph{dependent} SiDB is selected. It is an arbitrarily chosen SiDB $k$ (but not among the $d$ pre-determined negatively-charged SiDBs) whose charge state is always deduced by the charge states of all other SiDBs in the layout due to the population stability constraint. In other words, when enumerating all electron distributions for the remaining $n - d$ SiDBs in the layout, we can always stop at iteration $2^{n - 1 - d}$ (or $3^{n - 1 - d}$), because the final SiDB's charge state is inherently determined by the charge states of all others.
	Hence, the charge state of SiDB $k$ can be described depending on $n_{i}$, $i \neq k$ as follows:
	
	\begin{align}
		\centering
		\mathit{SiDB_{k,-}} &\iff \mu_{-} + \underbrace{\sum_{i, i \neq k} V_{k,i}(n_{i})}_{V_{\text{local},k}} \cdot \, q_{e} < 0 \\
		\mathit{SiDB_{k,+}} &\iff \mu_{+} + \sum_{i, i \neq k} V_{k,i} \cdot q_{e} > 0 \\
		\mathit{SiDB_{k,0}} &\iff \text{otherwise}
	\end{align}

	Thereby, it becomes non-imperative to enumerate the charge state of the $k$-th~SiDB, which reduces the search space by another factor of $2$ in \mbox{2-state} (or $3$ in \mbox{3-state}) simulation. 
	
	Hence, via the introducing of the two \emph{Physically-informed Search Space Pruning} concepts (detecting negatively charged SiDBs to satisfy population stability and introducing the \emph{dependent} SiDB) mentioned above, the search space is reduced by a factor of $2^{d+1}$ (or $3^{d+1}$) without losing exactness.
	
	\begin{algorithm}[!t]
		\SetAlgoLined
		\DontPrintSemicolon
		\KwIn{SiDB layout $\mathit{L}$ comprised of $n$ dangling bonds}
		\KwIn{Physical parameters $P = \{ \mu_{-}, \mu_{+}, \lambda_{\mathit{tf}}, \epsilon_{r} \}$}
		\KwOut{A set $S$ of SiDBs that must be negatively charged} 
		
		Electron distribution $D \gets [D_1 = -1, \dots, D_n = -1]$ \label{line:negative_sidbs:negative_distribution} \;
		$V_{\mathit{local},i} \gets \sum_{j, j \neq i} V_{i,j}$ given $D$\tcp*{\refequa{local_potential}}
		$S \gets \emptyset$ \;
		\ForEach{$\mathit{db} \in L$\label{line:negative_sidbs:looping_through_changed_sidbs}}
		{
			\If{$\mu_{-} + V_{\mathit{local},\mathit{db}} \cdot q_{e} < 0$\label{line:negative_sidbs:detected}}
			{
				$S \gets S \cup \{\mathit{db}\}$ \;
			}
		}
		\Return $S$ 	\label{line:negative_sidbs:return} 
		\caption{Negative SiDB Detection}
		\label{algo:negative_sidbs} 
	\end{algorithm}
	
	\subsubsection{Partial Solution Caching} \label{sec:quickexact:caching}
	
	As mentioned before, it is \mbox{runtime-expensive} to recompute the local electrostatic potential at each SiDB position every time the charge distribution changes.
	To reduce the runtime of this calculation step, \emph{Partial Solution Caching} is introduced. This means that the entire calculation according to \refequa{local_potential} is not carried out every time the charge distribution changes. Instead, the previous local electrostatic potentials are reused and updated while only considering the SiDBs at which the charge state has changed. 
	
	This can be conducted as summarized in the pseudocode shown in \refalgo{potential_caching}. The SiDB layout~$L$, the physical parameters~$P$ as well as the original and updated charge distributions,~$D$ and~$D'$, respectively, are used as the algorithm's input. As a first step, all SiDBs that have undergone a change in their charge state are identified and collected as~$C$~(\refline{potential_caching:changed_sidbs}). Subsequently, the local electrostatic potentials of the SiDBs that have remained unchanged in their charge state are updated~(\refline{potential_caching:updating_potential}). Thus, the number of SiDBs that have changed in their charge state (\ie, the number of elements in $\mathit{L \cap C}$) determine the number of iteration steps required to update the local electrostatic potentials. 
	
	
	
	\begin{algorithm}[!t]
		\SetAlgoLined
		\DontPrintSemicolon
		\KwIn{SiDB layout $\mathit{L}$ comprised of $n$ dangling bonds}
		\KwIn{Physical parameters $P = \{ \mu_{-}, \mu_{+}, \lambda_{\mathit{tf}}, \epsilon_{r} \}$}
		\KwIn{Original electron distribution $D$}
		\KwIn{Updated electron distribution $D'$}
		$C \gets D' \setminus D$ \label{line:potential_caching:changed_sidbs}\;
		\ForEach{$\mathit{db}' \in L \cap C$\label{line:potential_caching:looping_through_changed_sidbs}}
		{
			\ForEach{$\mathit{db} \in L \setminus (L \cap C)$\label{line:potential_caching:foreach_sidb}}
			{
				$V_{\mathit{local}, \mathit{db}} \gets V_{\mathit{local}, \mathit{db}} + V_{\mathit{db},\mathit{db}'} \cdot C[\mathit{db}']$ given $P$\label{line:potential_caching:updating_potential}
			}
		}
		\caption{Update Potential Cache}
		\label{algo:potential_caching} 
		
	\end{algorithm}

	
	\subsubsection{Effective State Enumeration} \label{sec:quickexact:gray_code}
	The previously presented technique, \emph{Partial Solution Caching}, is the most effective when the number of SiDBs that are changing their charge state between each iteration is small (\ie, $|\mathit{L \cap C}|$ is small). This third technique focuses on the effective iteration of electron distributions such that the least amount of SiDBs change their charge state in each step. 
	
	Electron distributions $D$ can be interpreted as $b$-ary digit strings of length $n$, where $b \in \{2, 3\}$ is the \emph{base}, \ie, the number of SiDB states.\footnote{For the sake of simplicity, we assume $b = 2$ in the following. However, the proposed technique works for arbitrary values of $b$.} To this end, value $v$ at position $D[i]$ indicates that SiDB $i$ is assigned charge state $v$. It follows that if the \emph{Hamming distance} of $D$ to $D'$ (from one iteration step to the next) is minimal, the number of SiDBs that have changed their state is also minimized. It is further known that the average Hamming distance between increments of binary strings converges to $2$, \ie, when simply iterating through electron distributions in binary order, $2$ SiDBs change their polarity in each step on average. By using \emph{Gray code} as the enumeration scheme, we can reduce this number to $1$, since in Gray code order, only a single digit changes its value from one step to the next.
	This approach allows for maximally efficient updating of the local electrostatic potential (as described in \refalgo{potential_caching}), since only one charge state in the charge distribution changes at a time~($|\mathit{L \cap C}|=1$), reducing computational cost and improving efficiency by roughly another factor of $2$.
	
	
	\subsection{Resulting Performance Improvements} \label{sec:quickexact:complexity}
	
	The introduced optimization techniques elaborated on in the previous section, improve the performance of \emph{exact} physical simulation by several orders of magnitude. For \mbox{2-state} simulation, the total theoretical performance gain given as the achieved search space reduction is composed as follows:
	\begin{equation*}
		\underbrace{2^{d+1}}_{\text{\refsec{quickexact:pruning}}} \cdot \underbrace{\frac{n}{|L \cap C|}}_{\text{\refsec{quickexact:caching}}} \cdot \underbrace{2}_{\text{\refsec{quickexact:gray_code}}} \approx 2^{d+1} \cdot n,
	\end{equation*}
	since $|L \cap C| \approx 2$. 
	\begin{example}
		Consider an exact \mbox{2-state} simulation of a layout consisting of $30$~SiDBs \mbox{($n = 30$)}, where five SiDBs (\mbox{$d=5$}) are detected to be negatively charged in a physically valid charge configuration. While the state-of-the-art simulator \exgs would require $2^{30}$ simulation steps, using the techniques introduced above allows to conduct the simulations approximately \mbox{$2^{5+1} \cdot 30 \approx 2000$} times faster.
	\end{example}
	
	\subsection{Implementation Details} \label{sec:quickexact:implementation}
	
	\begin{algorithm}[!t]
		\SetAlgoLined
		\DontPrintSemicolon
		\KwIn{SiDB layout $\mathit{L}$ comprised of $n$ dangling bonds}
		\KwIn{Physical parameters $P = \{ \mu_{-}, \mu_{+}, \lambda_{\mathit{tf}}, \epsilon_{r} \}$}
		\KwOut{All valid electron distributions with their respective energies}
		$\mathit{DE} \gets \emptyset$ \;
		$S \gets \textsc{Negative SiDB Detection}(L, P)$ \label{line:physical_simulation:negative_sidbs}\tcp*{\refalgo{negative_sidbs}}
		$L_S \gets L \setminus S$ \label{line:physical_simulation:remaining_sidbs}\;
		dependent SiDB $\tilde{\mathit{db}} \gets {L_S}[1]$ \;
		
		Electron distribution $D \gets [D_1 = -1, \dots, D_n = -1]$ \; 
		$V_{\mathit{local},i} \gets \sum_{j, j \neq i} V_{i,j}$ given $D$\tcp*{\refequa{local_potential}}
		
		\ForEach{possible $D'$ of $L_S \setminus \{\tilde{\mathit{db}}\}$ \text{in Gray code order}}
		{			
			$\textsc{Update Potential Cache}(L, P, D, D'$)\tcp*{\refalgo{potential_caching}}\label{line:physical_simulation:caching}
			\If{$D' \text{ is phys. valid given } P$\label{line:physical_simulation:valid}}
			{
				$E \gets \text{energy of } L \text{ given } D' \text{ and } P$\tcp*{\refequa{energy}}
				$\mathit{DE} \gets \mathit{DE} \cup \{(D', E)\}$ \label{line:physical_simulation:collection}\;
			}
			
			$D \gets D'$ \;
		}
		
		\Return $\mathit{DE}$ \label{line:physical_simulation:return} \; 
		\caption{\quickexact}
		\label{algo:physical_simulation}
	\end{algorithm}
	
	The techniques proposed above can be implemented on top of any combinatorial exact physical simulator for SiDBs. In this work, we took \exgs as a basis and realized the proposed concepts on top of it. This led to the efficient physical simulator \quickexact. How \quickexact works as a whole is sketched in  \refalgo{physical_simulation}. The algorithm starts with an SiDB layout comprised of $n$ SiDBs and the physical parameters~$P$ as input. First, all SiDBs that must be negatively charged in a physically valid charge distribution are detected with \refalgo{negative_sidbs} as explained before (\refline{physical_simulation:negative_sidbs}). All remaining SiDBs (\ie, SiDBs that can also be neutrally charged) are collected as $L_{S}$ in \refline{physical_simulation:remaining_sidbs}. Then, the first SiDB of $L_{S}$ is arbitrarily chosen as the \emph{dependent} SiDB $\tilde{\mathit{db}}$. Subsequently, all possible charge distributions of the SiDBs in $L_S \setminus \{\tilde{\mathit{db}}\}$ are enumerated in \emph{Gray code order}. 
	To reiterate, this is an efficient way to cover all possible charge distributions by changing the charge state of only one SiDB in each iteration. The local electrostatic potential of the current charge distribution cache is updated with \refalgo{potential_caching} (\refline{physical_simulation:caching}) to check if the physical validity is fulfilled (\refline{physical_simulation:valid}). If the charge distribution under consideration is physically valid, it is stored together with its corresponding electrostatic potential energy~(\refline{physical_simulation:collection}).
	
Ultimately, all physically valid charge distributions with their corresponding electrostatic potential energy are returned~(\refline{physical_simulation:return}).

	\section{Experimental Evaluations} \label{sec:eval}
	
	To demonstrate the applicability and improved runtime of the resulting tool \quickexact in comparison to the state of the art, an exhaustive experimental evaluation is conducted, whose results are presented in this section. To this end, first the experimental setup is described in \refsec{eval:setup}. Afterwards, the respectively obtained experimental results are presented in \refsec{eval:random} considering two different classes of SiDB layouts, namely \begin{enumerate*}
		\item randomly-generated,
		and
		\item established ones from the literature.
	\end{enumerate*}
	Finally, the 
	obtained results are discussed in \refsec{eval:discussion}.

	\subsection{Experimental Setup} \label{sec:eval:setup}
	
	\quickexact has been implemented as summarized in \refsec{quickexact:implementation}. We compare the performance of \quickexact with \exgs. All code has been implemented in C++ on top of the \fiction\footnote{\url{https://github.com/cda-tum/fiction}} framework~\cite{fiction} which is part of the \mbox{\emph{Munich Nanotech Toolkit}~(MNT)} and compiled with AppleClang 14.0.0. All experiments were carried out on a macOS 13.0 machine with an Apple Silicon M1 Pro SoC with \SI{32}{\giga\byte} of integrated main memory.

	\subsection{Considered Layouts and Obtained Results} \label{sec:eval:random}
	
	To validate the efficiency and accuracy of physical simulation, it is important to use a wide variety of SiDB layouts. To this end, in the first series of evaluations, 280 \mbox{randomly-generated} layouts with different numbers of SiDBs per layout (ranging from $20$ to $33$ SiDBs per layout) are used. In a second series of evaluations, all $12$ gate layouts from the \emph{Bestagon} gate library~\cite{walter2022hexagons} are simulated. The Bestagon gate library is a universal collection of hexagonal SiDB standard tiles for circuit layout design that have been created using Reinforcement Learning with a focus on realistic design constraints imposed by fabrication capabilities. 
	
	\begin{table}[!t]
		\centering
		\caption{Exact simulation of randomly generated layouts}
		\label{tab:randomly_generated_layouts}
		\begin{minipage}{0.65\linewidth}
			\centering
			\begin{adjustbox}{max width=\linewidth}
				\begin{tabular}{
						c 
						S[table-format=2.0] 
						c 
						S[table-format=5.2] 
						c 
						S[table-format=2.2] 
						}
					\toprule
					\multicolumn{2}{c}{\multirow{2}{*}{\textsc{Benchmark}}} &  \phantom{m} & \multicolumn{3}{c}{\multirow{2}{*}{\textsc{Physical Simulation}\footnote{All runtime values are in seconds}}} \\
					\multicolumn{6}{c}{~}  \\
					\cmidrule(lr){1-2} \cmidrule(lr){4-6}
					\multicolumn{1}{c}{\multirow{2}{*}{\#SiDBs}} & \multicolumn{1}{c}{\multirow{2}{*}{\#inst.}}  & &  \multicolumn{1}{c}{\multirow{2}{*}{\exgs~\cite{ng2020thes}}} && \multicolumn{1}{c}{\multirow{2}{*}{\emph{QuickExact}}}
					\\
					\multicolumn{6}{c}{~}  \\
					\midrule
					\multirow{1}{*}{}   
					 $33$ & 20 && > 3600  && 75.67 \\ 
					\addlinespace[0.5em]
					\multirow{1}{*}{}   
					 $32$ & 20 && > 3600  && 50.69 \\ 
					\addlinespace[0.5em]
					\multirow{1}{*}{}   
					 $31$ & 20 && > 3600  && 11.42 \\ 
					\addlinespace[0.5em]
					\multirow{1}{*}{}   
					 $30$ & 20 && > 3600 && 3.24 \\ 
					\addlinespace[0.5em]
					\multirow{1}{*}{}   
					$29$  & 20 && > 3600  && 2.83 \\ 
					\addlinespace[0.5em]
					\multirow{1}{*}{}             
					$28$ & 20 && > 3600 && 2.95 \\ 
					\midrule
					\multirow{1}{*}{}   
					 $27$ & 20 &&  3408.25  && 0.98\\ 
					\addlinespace[0.5em]
					\multirow{1}{*}{}   
					$26$  & 20 && 1700.68   && 0.28\\
					\addlinespace[0.5em]
					\multirow{1}{*}{}   
					 $25$ & 20 && 827.67   && 0.11\\
					\addlinespace[0.5em]
					\multirow{1}{*}{}       
					$24$  & 20 && 414.38 && 0.05\\ 
					\addlinespace[0.5em]
					\multirow{1}{*}{}   
					$23$ & 20 && 207.25  && 0.04 
					\\ \addlinespace[0.5em]
					\multirow{1}{*}{}   
					 $22$  & 20 && 103.64  && 0.00\\ 
					\addlinespace[0.5em]
					\multirow{1}{*}{}   
					$21$  & 20 && 48.60 && 0.00\\ 
					\addlinespace[0.5em]
					\multirow{1}{*}{}   
					 $20$  & 20 && 25.34 && 0.00\\ 
					\midrule
					\multirow{1}{*}{\emph{Sum}\footnote{without $>$\SI{3600}{} rows}}
					&& & 6735.81 && 1.49 \\  \addlinespace[0.5em]
					\multirow{1}{*}{\emph{\%}}
					&& & 100  && 0.02 \\ 
					\bottomrule
				\end{tabular}
			\end{adjustbox}
		\end{minipage}
	\end{table}

	\begin{table}[!t]
		\centering
		\caption{Exact simulation of established gate layouts}
		\label{tab:established_gates}
		\begin{minipage}{0.9\linewidth}
			\centering
			\begin{adjustbox}{max width=\linewidth}
				\begin{tabular}{
						l 
						S[table-format=2.0] 
						S[table-format=1.0] 
						c 
						S[table-format=4.2] 
						c 
						S[table-format=1.2] 
						}
					\toprule
					\multicolumn{3}{c}{\multirow{2}{*}{\textsc{Benchmark\cite{walter2022hexagons}}}} &  \phantom{m} & \multicolumn{3}{c}{\multirow{2}{*}{\textsc{Physical Simulation}\footnote{All runtime values are in seconds}}} \\
					\multicolumn{7}{c}{~}  \\
					\cmidrule(lr){1-3} \cmidrule(lr){5-7}
					\multicolumn{1}{c}{\multirow{2}{*}{Name}} & \multicolumn{1}{c}{\multirow{2}{*}{\#SiDBs}} & \multicolumn{1}{c}{\multirow{2}{*}{\#inst.}}  & &  \multicolumn{1}{c}{\multirow{2}{*}{\exgs~\cite{ng2020thes}}} && \multicolumn{1}{c}{\multirow{2}{*}{\emph{QuickExact}}}
					\\
					\multicolumn{7}{c}{~}  \\
					\midrule
					\multirow{1}{*}{DOUBLE WIRE}   
					&  28 & 4 && 1211.85 && 0.48\\ \addlinespace[0.5em]
					\multirow{1}{*}{CX}   
					& 27  & 4 && 599.61 && 0.28\\ \addlinespace[0.5em]
					\multirow{1}{*}{HA}   
					&  24  & 4&& 71.35  && 0.04\\ \addlinespace[0.5em]
					\multirow{1}{*}{AND}   
					& 21 &4 && 58.53   && 0.06
					\\ \addlinespace[0.5em]
					\multirow{1}{*}{XOR}       
					& 21  & 4 && 8.40   && 0.00\\ \addlinespace[0.5em]
					\multirow{1}{*}{OR}   
					&  21 & 4 && 6.84   && 0.00\\ \addlinespace[0.5em]
					\multirow{1}{*}{XNOR}   
					& 21  & 4 && 8.41   && 0.00\\ \addlinespace[0.5em]
					\multirow{1}{*}{FO2}   
					&  20 & 2 &&  2.16  && 0.00\\ \addlinespace[0.5em]
					\multirow{1}{*}{NOR}             
					& 19 & 4 &&  2.07  && 0.00\\ \addlinespace[0.5em]
					\multirow{1}{*}{NAND}   
					& 19  & 4 &&  2.07   && 0.00\\ \addlinespace[0.5em]
					\multirow{1}{*}{INV}   
					&  18 & 4 && 1.04  && 0.00\\  \addlinespace[0.5em]
					\multirow{1}{*}{WIRE}   
					&  15 & 4 &&  0.26  && 0.00\\  \midrule
					\multirow{1}{*}{\emph{Sum}}
					& && & 1972.58  && 0.91 \\  
					\addlinespace[0.5em]
					\multirow{1}{*}{\emph{\%}}
					& && & 100  && 0.05\\ 
					\bottomrule
				\end{tabular}
			\end{adjustbox}
		\end{minipage}
	\end{table}
	
	All obtained results are summarized in \reftab{randomly_generated_layouts} and \reftab{established_gates} for the \mbox{randomly-generated} layouts and the established \emph{Bestagon} library, respectively. In \reftab{randomly_generated_layouts}, the first two columns provide the number of SiDBs and the number of simulated layouts (instance count), respectively. Subsequently, the simulation runtimes of \exgs and \quickexact are stated. \reftab{established_gates} has the same structure; however, the name of the investigated gate is added as first column. Since each 2-input gate can be simulated in up to four configurations, with input perturbers specifying one of the possible $2^{4}$ input combinations, the instance count is generally $4$. However, the fan-out tile~(FO2) is a 1-input/2-output gate and thus only provides $2$ instances. WIRE and INV, despite being 1-input tiles, have instance count~$4$ because two different types---straight and diagonal---exist for each.
	
	
	\subsection{Discussion} \label{sec:eval:discussion}
	The simulation of the \mbox{randomly-generated} layouts and the established \emph{Bestagon} library reveals a massive performance advantage of \quickexact compared to \exgs. For \mbox{randomly-generated} layouts, \exgs  yields a timeout (one hour) for $120$ layouts ($40$ of them need even more than $24$~hours), while \quickexact determines valid solutions to all instances within seconds, resulting in a total runtime advantage of a factor of $5000$ compared to the state of the art,~\exgs.
	
	These results are again confirmed by the simulation of the \emph{Bestagon} gate library. While \exgs takes about \SI{2000}{\second} to simulate all gates in all configurations, \quickexact takes less than \SI{1}{\second}, again a reduction in runtime of more than a factor of~$2000$.
	
	This runtime advantage in all cases demonstrates the clear benefit that the techniques proposed in this work have on exact physical simulation of SiDBs. 
	We have hereby enabled the obtainment of exact simulation results for large layouts that were previously impossible to process.

	\section{Conclusions} \label{sec:concl}

	In this work, a novel approach, \quickexact, for exact physical simulation of SiDB layouts was presented with a special focus on runtime improvement. To this end, three techniques were proposed: \begin{enumerate*}
		\item \mbox{Physically-informed} Search Space Pruning,
		\item Partial Solution Caching, and
		\item Effective State Enumeration.
		\end{enumerate*}
	In an extensive experimental evaluation, these techniques lead to runtime improvements of up to a factor of $5000$ compared to the state of the art, \exgs, as shown on two different sets of SiDB layouts: \begin{enumerate*}
		\item \mbox{randomly-generated} ones, and
		\item established standard gate tiles from the literature.
	\end{enumerate*}
	In case of random layouts, \exgs did not yield any results within one hour for $120$ of the considered $280$~layouts ($40$ of them need even more than $24$~hours), while \quickexact required \SI{75.67}{\second} at maximum. In case of the standard gate tiles, \exgs needed a total of around \SI{2000}{\second} while \quickexact stayed below \SI{1}{\second}, emphasizing its immense performance advantage.  
	To support open research and open data, we plan to make the implementation of \quickexact, all evaluation data, and the test scripts publicly available in case of a positive evaluation of this work.
	
	\balance
	\bibliographystyle{IEEEtran}
	\bibliography{./bib/IEEEabrv.bib, ./bib/Bibliography.bib}
	
\end{document}